\documentclass[11pt]{article}

\usepackage{amsmath}
\usepackage{amscd}
\usepackage{amssymb}
\usepackage{amsfonts}
\usepackage{eufrak}
\usepackage[mathscr]{euscript}
\usepackage{pstricks}
\usepackage{pst-node}
\usepackage[dvips]{pstcol}\textheight=8.5truein
\def \PBz{${{\mathrm{B}^0}}$}
\def \PaBz{${\mathrm{\overline{B}^0}}$}
\def \PBzL{${{\mathrm{B_L}}}$}
\def \PBzH{${{\mathrm{B_H}}}$}
\def \PDp{${\mathrm{D^+}}$}
\def \PDm{${\mathrm{D^-}}$}
\def \Pgpp{${\mathrm{\pi^+}}$}
\def \Pgpm{${\mathrm{\pi^-}}$}
\def \PJ{${\mathrm{J /\psi}}$}
\def \PK{${\mathrm{K}}$}

\begin{document}

\title{\bf An open quantum system approach to the B-mesons system}
\author{Raffaele Romano\footnote{E-mail address:
	rromano@ts.infn.it} 
	\\ Dipartimento di Fisica Teorica \\ Universit\`a di Trieste \\ Italy}
\date{Jaca, Spain, May 2001}
\maketitle

\begin{abstract}

\noindent
In this talk we consider a non standard evolution for the \PBz-\PaBz 
system, namely, an evolution in the open quantum systems framework.
Such approach is justified by the very high sensitivity of experiments
studying CP-violating phenomena in the B-mesons sector, very near to
the one required to test some possible scenarios induced by quantum
gravity at the Planck scale, whose effects at low energy can be
described by a heat bath.  
We adopt a phenomenological point of view, introducing six new
parameters that fully describe this kind of evolution without
referring to a specific model for the microscopic interaction.
We outline the main differences between this approach and the usual
one in the description of evolution and decay of single mesons
or correlated pairs.

\end{abstract}


\section{Introduction}

The physics of b-flavoured hadrons is at present one of the most promising
fields testing Standard Model predictions concerning some
interesting physical processes like CP-violating phenomena and
flavour oscillations.
Presently, several experiments are collecting data; other very accurate 
experiments are planned to start in the near future. In this context
it is also possible to study some exotic scenarios of physics beyond the
Standard Model; in this contribution we consider an open quantum
system treatment for the \PBz-\PaBz system~\cite{BFR} producing
decoherence\footnote{For a preliminary analysis see~\cite{BF}; for a
different approach to decoherence in the B-mesons system see~\cite{GB}}.

A wide variety of physical systems has been studied in this
framework. The formalism of open quantum systems, initially developed 
in quantum optics~\cite{qo1,qo2,qo3}, has been also applied to study 
phenomena of quantum relaxation in magnetic resonance~\cite{slic},
to foundational aspects of quantum mechanics~\cite{fond} 
and, lately, to model some typical situations in
particle physics: neutral mesons systems~\cite{mes1,mes2},
neutrinos~\cite{neu}, photons~\cite{phot}, neutrons
interferometry~\cite{nei}.

An open system~\cite{os1,os2,os3} 
is a system S that interacts with its external environment E, and, then,
exchanging energy and entropy with it. Whereas the dynamics of the
composite system T~=~S~+~E is unitary, this is no longer true for the 
dynamics of the system S alone, obtained tracing the time-evolution of
T over the degrees of freedom of the environment E. In general, the
resulting reduced dynamics is complicated by memory effects. It is
possible to get a Markovian evolution for the system S (that is, a
memoryless dynamics) if essentially two assumptions hold: the
interaction between S and E must be weak, and there must be no initial 
correlation between them~\cite{os1,os2,os3}.

Since we are interested in time-evolutions causing decoherence, that
is transitions from pure states to mixtures, we must describe the
state of S by means of statistical operators $\rho$, that is
Hermitian and positive trace-class matrices with unit trace. The
time-evolution for $\rho$ is given by

\begin{equation}
\dot\rho(t) = L[\rho(t)] = -i\,H\,\rho(t)\, +\, 
i\,\rho(t)\,H^\dagger\,+\,L_D [\rho(t)],
\label{evolution}
\end{equation}

\noindent
where $L$ is the generator of the dynamics,
$H$ is the Weisskopf-Wigner Hamiltonian describing the system S
(non-Hermitian because the system is unstable) and $L_D$ is the 
generator of the dissipative part, emboding the interaction
with the environment. The form of $L_D$, univoquely fixed by the
Kossakowski theorem~\cite{os1,os2},is very general since it
characterizes any Markovian dynamics:

\begin{equation}
L_D[\rho(t)] = -\frac{1}{2}\Bigl\{\sum_i A_i^{\dagger}A_i,\rho(t)\Bigr\}+
\sum_i A_i\rho(t)A_i^{\dagger}\, ,
\label{lindblad}
\end{equation}

\noindent
where the set of operators $A_i$ depends on the details of the interaction.

The time-evolution maps $\rho(0) \rightarrow \rho(t)=\Lambda_t
[\rho(0)]$, where $\Lambda_t = e^{Lt}$, satisfy the forward in time
composition law: $\Lambda_{t+s}=\Lambda_t \circ \Lambda_s$, with
$t,s\geq 0$, characteristic of irreversible phenomena; furthermore,
they are completely positive~\cite{os1,os2}. 
This last property is noteworthy in particle physics, because
it guarantees the preservation of positivity of evolving density
matrices describing entangled systems. Indeed, simple positivity is
not sufficient for a proper statistical interpretation of the
formalism of density matrices when they represent entangled 
systems~\cite{os1,os2} (counter examples are found in~\cite{mes3}). 
Moreover, the evolution maps $\Lambda_t$ allow transitions from pure
to mixed states. 

The entropy of open systems can increase as well as decrease;
nevertheless, in what follows, we will consider only entropy
increasing time-evolutions because we are interested in reduced
dynamics producing states less ordered as time passes.

As we shall shortly see in the next section, the phenomenonogical
approach to the \PBz-\PaBz system is usually performed in the two
dimensional Hilbert space of flavours; then we shall consider $2
\times 2$ density matrices $\rho$ that can be expanded over
the Pauli matrices $\sigma_1$ ($i = 1,2,3$) and the $2 \times 2$
identity $\sigma_0$:  

\begin{equation}
\rho=\sum_{i=0}^{3}\rho_i \, \sigma_{i} \, .
\label{rhopauli}
\end{equation}

It is possible to give a vectorial representation of the density
matrix $\rho$, namely, defining a $4$-vector $\vert \rho \rangle$
whose components are the real coefficients $\rho_i$
in~(\ref{rhopauli}):

\begin{equation*}
\vert \rho \rangle \equiv \left(
\begin{array}{l}
\rho_0 \\ 
\rho_1 \\ 
\rho_2 \\ 
\rho_3
\end{array}
\right)
\end{equation*}

accordingly, since equation~(\ref{evolution}) is
linear in time, it can be rewritten as a Schr{\"o}dinger-like equation:

\begin{equation}
\partial_t \vert \rho(t) \rangle \,=\,({\cal H}\,+
{\cal D})\,\vert \rho(t) \rangle\, ,
\label{evolution2}
\end{equation}

\noindent
where $\cal H$ and $\cal D$ are $4 \times 4$ real matrices, the
Hamiltonian and the dissipative part respectively. The non-standard
part of the dynamics is embodied in the symmetric matrix {\cal D}:

\begin{equation}
{\cal D}=-2 \begin{pmatrix}
0 & 0 & 0 & 0 \cr
0 & a & b & c \cr
0 & b & \alpha & \beta \cr
0 & c & \beta & \gamma
\end{pmatrix}\, ,
\label{dissipative}
\end{equation}

\noindent
with $a,b,c,\alpha,\beta,\gamma$ six real phenomenological constants,
constrained by the inequalities, necessary and sufficient to Complete
Positivity~\cite{os1}:

\begin{equation}
\begin{matrix}
&2R \equiv \alpha+\gamma-a\,\geq\,0\, ,
\, RS\geq b^2 ,\cr
&2S \equiv a+\gamma-\alpha\,\geq\,0\, ,
\, RT\geq c^2 ,\cr
&2T \equiv a+\alpha-\gamma\,\geq\,0\, ,
\, ST\geq \beta^2 ,\cr
&RST\,\geq\,2bc\beta+R\beta^2+S c^2+T b^2 .
\end{matrix}
\end{equation}

Since these coefficients are small\footnote{A rough evaluation of
their magnitude can be done: ${\cal D}_{ij}\sim
\epsilon^{2}_{S}/\epsilon_{E}$, where $\epsilon_{S}$ and
$\epsilon_{E}$ are the characteristic energies of the system and of
the environment respectively} with respect to the entries of 
$\cal H$, it is possible to solve~(\ref{evolution2}) up to first order
in these parameters: 

\begin{eqnarray}
\vert \rho(t) \rangle = e^{{\cal H} t} \vert \rho (0) \rangle +
\int_{0}^{t} ds \, e^{{\cal H} (t-s)} {\cal D}\, e^{{\cal H} s} \vert
\rho(0) \rangle\, .
\label{solution}
\end{eqnarray}

The first term describes the standard evolution of the density matrix,
the second one accounts for the interaction with the environment. 

In the following we won't report the explicit form of statistical
operators at time $t$ because it is not physically relevant.
The computations have been performed in vectorial
representation and using the approximation~(\ref{solution}).


\section{Decay of neutral B-mesons}

The neutral B-mesons system can be represented by statistical
operators $\rho$ acting on the two dimensional flavour Hilbert space,
spanned by the physical states $\vert$\PBz$\rangle$ and
$\vert$\PaBz$\rangle$. In a dissipative context, their time-evolution
in vectorial representation is given by~(\ref{solution}); the
computation has been performed working in the basis of the 
eigenstates of the Hamiltonian $H$, denoted by $\vert$\PBzL$\rangle$
and $\vert$\PBzH$\rangle$, where the subscripts mean ``light'' and
``heavy'' respectively:

\begin{equation}
\label{cicciolina}
\left\{ \begin{array}{l}
H \vert {\rm B_H} \rangle = \lambda_H \vert {\rm B_H} \rangle, \\
H \vert {\rm B_L} \rangle = \lambda_L \vert {\rm B_H} \rangle,
\end{array}
\right.,
\quad \lambda_{H,L} = m_{H,L} - \frac{i}{2} \gamma_{H,L}
\end{equation}

\noindent
where $m_{H,L}$ and $\gamma_{H,L}$ are the masses, respectively the
widths, of the corresponding eigenstates.

This choice of basis provides an invariant formalism under
independent phase redefinitions of the two basis vectors. 
Therefore it makes no commitment on the magnitude of CP (and CPT)
violating effects.  

In the density matrix formalism, the mean value of an observable 
$\cal O$ is obtained from:

\begin{equation}
\langle {\cal O}(t) \rangle = Tr\left[{\cal O} \rho (t)\right].
\label{meanvalue}
\end{equation}

\noindent
Then, the probability rates for the decays of \PBz and \PaBz in
different channels, \PBz, \PaBz $\rightarrow f$, can be written as

\begin{center}
${\cal P}_f(B^0,t)$ = Tr$[{\cal O}_f \rho(t)]$, \,
with $\rho(0)=\vert$\PBz$\rangle \langle$\PBz$\vert$;

${\cal P}_f(\overline{B}^0,t)$ = Tr$[{\cal O}_f \rho(t)]$, \,
with $\rho(0)=\vert$\PaBz$\rangle \langle$\PaBz$\vert$,
\end{center}

\noindent
where the operator ${\cal O}_f$ is defined by: 

\begin{equation}
{\cal O}_f = \begin{pmatrix} 
o_1&o_3\cr o_4&o_2
\end{pmatrix},
\label{obs}
\end{equation}

\noindent
with entries:

\begin{center}
$o_1 = \vert{\cal A}($\PBz$ \rightarrow f)\vert^{2},\,\,\, 
o_3 = [{\cal A}($\PBz$ \rightarrow f)]^*  
{\cal A}($\PaBz$ \rightarrow f),$ \\
$o_2 = \vert{\cal A}($\PaBz$ \rightarrow f)\vert^{2}, \,\,\,
o_4 =[{\cal A}($\PaBz$ \rightarrow f)]^*
{\cal A}($\PBz$ \rightarrow f)$,
\end{center}

\noindent
where $\cal A$ the the decay amplitude for the considered process.

The standard dynamics is characterized by $\Delta m~=~m_H~-~m_L$,
$\Delta~\Gamma~=~\gamma_L~-~\gamma_H$ and $\Gamma~=~\gamma_L +
\gamma_H /2$; it is customary to renormalize these parameters, and the
time variable, in the following way: 

\begin{equation}
\omega = \frac{\Delta m}{\Gamma},\quad \delta = \frac{\Delta \Gamma}{2\Gamma},
\quad \tau = t \, \Gamma\,.
\end{equation}

The dissipative contribution is instead parametrized by particular 
functions of $a, b, c, \alpha, \beta$ and $\gamma$, denoted by $A, B,
C$ and $D$; their expressions are reported in~\cite{BFR}.

\subsection{Decay of a single B-meson}

Typical decays, relevant for the study of CP (and CPT)
violation, are the so-called semileptonic decays, in which the final
state $f$ contains a hadron $h$, a lepton $l$ and a neutrino (or
anti-neutrino): \PBz (\PaBz)$\rightarrow h^-l^+ \nu\,(h^+ l^-\overline{\nu})$.

\noindent
It is customary to parametrize the amplitudes
in~(\ref{obs}) as follows~\cite{cp1,cp2}:

\begin{center}
${\cal A}($\PBz$\rightarrow h^-l^+\nu) = {\cal M}_h (1-y_h)$,

${\cal A}($\PaBz$\rightarrow h^+l^-\overline{\nu}) = {\cal M}^{*}_{h} 
(1+y^{*}_{h})$,

${\cal A}($\PBz$\rightarrow h^+l^-\overline{\nu}) = 
z_h\,{\cal A}($\PaBz$\rightarrow h^+l^-\overline{\nu})$,

${\cal A}($\PaBz$\rightarrow h^-l^+\nu) =
x_h\,{\cal A}($\PBz$\rightarrow h^-l^+\nu)$,
\end{center}

\noindent
where the parameters $x_h$ and $z_h$ measure violations of the
rule $\Delta B = \Delta Q$, satisfied in the Standard Model, and the
coefficient $y_h$ measures deviations from CPT invariance. 

For sake of simplicity, the results we show below are obtained 
assuming $x_h = z_h = y_h = 0$; more complete results are
collected in~\cite{BFR}. The probability rates present new dissipative terms
absent in the standard case; for example, in the case of the decay
\PBz $\rightarrow h^+ l^-\overline{\nu}$:

\begin{equation}
\begin{split}
{\cal P}_{h^{+}}(B^0,\tau) = \frac{\vert{\cal M}_h
\vert^2}{2}\,e^{-\tau} \Bigl\{& \cosh{\delta \tau} - \cos{\omega \tau}\,\,
e^{-(A - D)\tau} + \\ 
&+ \sinh{\delta \tau} \Bigl( \frac{D}{\delta}\, - \frac{4\,\omega}
{\delta^2 + \omega^2}\, {\rm Im}\,(C) \Bigr)
+ \sin{\omega \tau} \Bigl( \frac{4\,\delta}{\delta^2 + \omega^2}\, 
{\rm Im}\,(C)- {\rm Re}\,(B) \Bigr)\Bigr\}.
\label{probdecay}
\end{split}
\end{equation}

If dissipation is neglected ($A = B = C = D = 0$) we obtain the
standard expression of the probability rate: the second row 
of~(\ref{probdecay}) vanishes and the damping exponential in the
first row disappears.

From a fit of the time dependence of~(\ref{probdecay}) it should be
possible, at least in principle, to get the parameters $A$, $D$, ${\rm
Re} (B)$ and ${\rm Im}(C)$.

Usually, instead of analyzing directly the decay rates, some
particular asymmetries, constructed by means of these rates, are taken
into account. The following one, for example, is extensively studied:

\begin{equation}
{\cal A}_{\Delta m}(\tau)\equiv\frac
{[{\cal P}_{h^-}(B^0,\tau)-{\cal P}_{h^-}(\overline{B^0},\tau)]-
[{\cal P}_{h^+}(B^0,\tau)-{\cal P}_{h^+}(\overline{B^0},\tau)]}
{{\cal P}_{h^-}(B^0,\tau)+{\cal P}_{h^-}(\overline{B^0},\tau)+
{\cal P}_{h^+}(B^0,\tau)+{\cal P}_{h^+}(\overline{B^0},\tau)}.
\label{asdeltam}
\end{equation}

\noindent
Indeed, in the standard case, this quantity is used to fit the parameter
$\Delta m$ introduced previously; when we consider the non standard 
evolution it assumes the simple form:

\begin{equation}
{\cal A}_{\Delta m}(\tau) = e^{-A\,\tau}\cos{\omega \tau}+
{\rm Re}\,(B)\,\sin{\omega\tau}
\label{deltamdiss}
\end{equation}

\noindent
that reduces, in the standard case (i.e. $A = B = C = D = 0$) to
${\cal A}_{\Delta m}(\tau) = \cos{\omega \tau}$.
We see that a study of the time dependence of this quantity should
provede, in principle, the dissipative parameters $A$ and ${\rm
Re}(B)$. 

The evaluation of the parameter $\Delta m$ in the standard case is
also performed considering time integrated rates

\begin{equation}
{\cal P}_f (B) = \frac{1}{\Gamma} 
\int_{0}^{\infty} d\tau {\cal P}_f (B, \tau)
\label{proint}
\end{equation}

\noindent
and introducing the asymmetry ${\cal A}_{\Delta m}^{\prime}$, defined in
analogy to~(\ref{asdeltam}), with the time integrated rates; the
expression of this quantity in the dissipative context is: 

\begin{equation}
{\cal A}_{\Delta m}^{\prime} = \frac{1}{1+{\omega}^2}\left\{1+\omega\,
\Re \,(B) +\frac{1}{1+\omega ^2}\left[(\omega^2 -1)\,A 
- 2\omega^2 D\right]\right\}.
\label{deltamdiss2}
\end{equation}

\noindent
In this case, since the dependence on time is lost, it is more
difficult to constrain the parameters $A$, ${\rm Re}(B)$ and $D$.
This is a general problem for time-integrated quantities.

Another class of relevant decays of single B-mesons has
a final state that is a CP eigenstate, for example: $f$ = \PDp
\PDm, \Pgpp \Pgpm, \PJ \, \PK. In this case it is convenient to define 
a new asymmetry:

\begin{equation}
{\cal A}_f (t) \equiv \frac{{\cal P}_f (B^0,t) - 
{\cal P}_f (\overline{B}^0,t)}{{\cal P}_f (B^0,t) + 
{\cal P}_f (\overline{B}^0,t)}.
\label{asimmf}
\end{equation}

This quantity would be rigorously zero in absence of CP
violation; it is studied because it enables to fit the
parameter $\beta$ appearing in the CKM matrix~\cite{cp1,cp2}. 
With the non-standard
contributions to the time-evolution it takes the form:

\begin{equation}
\begin{split}
{\cal A}_{f}(t) =& \frac{2 \xi_f}{\omega} {\rm Im}\,(C) - \sin{\omega \tau}
\left(\frac{2 {\rm Im}\,(\lambda_f)}{1+\vert \lambda_f\vert^2} +
\frac{2\xi_f}{\omega} {\rm Re}\,(C)\right) + \\
& + \cos{\omega \tau}  \left(\frac{1-\vert \lambda_f \vert ^2}
{1+\vert \lambda_f \vert ^2}-\frac{2\xi_f}{\omega} {\rm Im}\,(C) \right),
\label{asimfdis}
\end{split}
\end{equation} 

\noindent
where $\lambda_f$ is a phase-independent coefficient introduced in
order to parameterize the neutral mesons decays~\cite{cp1,cp2}.
A fit of this asymmetry would provide information about
the parameter $C$.

\subsection{Decay of correlated B-mesons}

Of particular interest are the decays of pairs of entangled B-mesons,
produced at the so-called B-factories. As already stated, the property
of Complete Positivity plays a fundamental role in this case.

The entangled B-mesons are generated from the decay of the
$\Upsilon$(4S) resonance: $\Upsilon$(4S)$\rightarrow$\PBz \PaBz; the
initial state is
$\vert \psi_B \rangle = \frac{1}{\sqrt{2}}
\left( \vert\right.$\PBz$,
-p\rangle\otimes\vert$\PaBz$,p\rangle-\vert $\PaBz$,
-p\rangle\otimes\vert $\PBz$,\left.p\rangle\right)$, 
and the initial density matrix 
$\rho(0) = \vert \psi_B \rangle \langle \psi_B \vert$.
Its time-evolution is supposed to be factorizable:

\begin{equation}
\rho(t_1,t_2)=(\Lambda_{t_1} \otimes \Lambda_{t_2})\rho(0),
\label{timeev}
\end{equation}

\noindent
where the first meson has evolved up to a time $t_1$ and
the second up to $t_2$. The probability of decay for the
first meson in a final state $f_1$ at time $t_1$ and for the second
meson in $f_2$ at $t_2$ is given by

\begin{equation}
{\cal P}(f_1, t_1; f_2, t_2) = Tr\left[\left({\cal O}_{f_1}
\otimes{\cal O}_{f_2}\right)\rho(t_1, t_2)\right],
\label{meanvalue2}
\end{equation}

\noindent
where the operators ${\cal O}_f$ have been introduced in~(\ref{obs}).

In the experimental analysis single time distributions are usually
taken into account; writing $t_1 = t^\prime + t$ and $t_2 = t^\prime$
we express these probabilities by

\begin{eqnarray}
{\cal P}_{f_1, f_2}(t)= \int_{0}^{+\infty}d t^{\prime}\,
{\cal P}(f_1, t^{\prime} + t; f_2, t^{\prime})
\label{singtimepo}
\end{eqnarray}

\noindent
for positive $t$, i.e. when the first meson decays after the second
one, and by

\begin{eqnarray}
{\cal P}_{f_1, f_2}(-\vert t \vert)= \int_{0}^{+\infty}d t^{\prime}\,
{\cal P}(f_1, t^{\prime} +t; f_2, t^{\prime}) \,\theta(t^{\prime}-
\vert t \vert)
\label{singtimene}
\end{eqnarray}

\noindent
for negative $t$ (when the first meson decays before). These
definitions fulfill the logical request: ${\cal P}_{f_1, f_2}(-\vert
t\vert) = {\cal P}_{f_2, f_1}(\vert t\vert)$.

We have analyzed the behavior of some characteristic quantities of
the double semileptonic decays, i.e. those decays in which both mesons
undergo a semileptonic decay.The first quantity we consider is the
asymmetry:

\begin{eqnarray}
{\cal R}(t) \equiv \frac{{\cal P}_{h^+, h^+}(t)+{\cal P}_{h^-, h^-}(t)}
{{\cal P}_{h^+, h^-}(t)+{\cal P}_{h^-, h^+}(t)}.
\label{ras}
\end{eqnarray}

Its relevant property is that in the standard case ${\cal R}(0) = 0$,
because ${\cal P}_{f, f}(0) = 0$, whereas taking in account the
dissipative contributions it becomes

\begin{eqnarray}
{\cal R}(0)=\frac{1}{2} \left[ A+\frac{\omega}{1+\omega^2} \left(
\omega\, {\rm Im}\,(B) - {\rm Re}\,(B) \right) \right],
\label{rasdis}
\end{eqnarray}

\noindent
regardless of the assumptions made about the coefficients 
$x_h, z_h$ and $y_h$.
Then a non-zero value of ${\cal R}(0)$ clearly shows the
presence of dissipative phenomena.

\noindent
Another interesting asymmetry is expressed by

\begin{eqnarray}
{\cal A}_{\Delta m}(\tau) \equiv
\frac{\left[{\cal P}_{h^+,h^-}(\tau) + {\cal
P}_{h^-,h^+}(\tau)\right]-\left[{\cal P}_{h^+,h^+}(\tau) + {\cal
P}_{h^-,h^-}(\tau)\right]}
{{\cal P}_{h^+,h^+}(\tau)+{\cal P}_{h^-,h^-}(\tau)+
{\cal P}_{h^+,h^-}(\tau)+{\cal P}_{h^-,h^+}(\tau)}.
\label{adeltame}
\end{eqnarray}

It is analogous to the one defined in~(\ref{asdeltam}) for the single
meson system, because it allows, if there is not dissipation, the
determination of the parameter $\Delta m$. In our case the dependence
on the dissipative parameters is:

\begin{equation}
{\cal A}_{\Delta m}(\tau) = \frac{e^{-A\tau}}{1+A}
\cos{\omega \tau}+(\omega \cos{\omega \tau} + \sin{\omega \tau})\,
{\rm Re}\,(\frac{B}{1-i\,\omega}).
\label{adeltamed}
\end{equation}

In the above expression we have assumed $\delta = 0$, a good
approximation since $\Delta \Gamma \ll \Gamma$.From the
time-dependence of the contributions in~(\ref{adeltamed}) it should be
possible to evaluate the dissipative parameters $A$ and $B$.


\section{Conclusions}

We have treated the neutral B mesons system
in the framework of quantum dynamical semigroups. The formalism is very
general and can be applied to describe the time evolution of this system 
when it is subject to a weak interaction with an external
environment. For instance, quantum gravity could give a motivation for such an
approach, since in this case the space-time itself would act as an
effective environment for any physical system. 

The open system framework allows a rough evaluation 
of magnitude of the dissipative parameters; they scale at most as 
$m_{\bf B}^2/M_F$, where $m_{\bf B}$ is the meson mass while $M_F$ is
a large fundamental mass scale. Tipically, $M_F$ coincides with the Planck 
scale so that $m_{\bf B}^2/M_F \sim 10^{-18}~{\rm GeV}$. This value is very 
small; however, the sofistication of the dedicated {\bf B}-experiments,
both at colliders(CDF-II, HERA-B, BTeV, LHC-$b$) and {\bf B}-factories
(BaBar, Belle, CLEO-III), is so high that the sensitivity
required to probe the presence of non-vanishing dissipative parameters
can be reached in just a few years of data 
taking~\cite{ball,btev,baba}.
If these parameters should be found not different from zero, the
inequalities expressing the Complete Positivity could be tested.
Then, the neutral B mesons physics is potentially a good laboratory
to test basic properties of open systems dynamics.


\section*{Acknowledgments}

I would like to thank Prof. Arno Bohm and the other organizers for the
invitation to the workshop; I am also grateful to F. Benatti and
R. Floreanini for many helpful discussions.
 
\noindent


\end{document}